\documentstyle[12pt]{article}

\def\be{\begin{equation}}
\def\ee{\end{equation}}

\begin{document}

\title {WEAK AND MAGNETIC INELASTIC SCATTERING OF ANTINEUTRINOS ON 
ATOMIC ELECTRONS}

\author{S.A. Fayans, L.A. Mikaelyan, V.V. Sinev}
\date{}
\maketitle

\centerline {\small \it Russian Research Centre -`` Kurchatov Institute'',
Kurchatov Sq. 1,}
\centerline {\small \it 123182 Moscow, Russia}

\begin{abstract}
Neutrino scattering on electrons is considered as a tool for
laborato-ry searches for the neutrino magnetic moment. We study
inelastic $\bar\nu_ee^-$-scattering on electrons bound in the
germanium ($Z=32$) and iodine ($Z=53$) atoms for antineutrinos
generated in a nuclear reactor core and also in the
$^{90}$Sr--$^{90}$Y and $^{147}$Pm artificial sources.
Using the relativistic Hartree-Fock-Dirac model, we calculate both
the magnetic and weak scattering cross sections for the recoil
electron energy range of 1 to 100~keV where a higher sensitivity to
the neutrino magnetic moment could be achieved. Particular attention
is paid to the approxi-mate procedure which allows us to take into
account the effects of atomic binding on the inelastic scattering
spectra in a simple way.
\end{abstract}

\vspace{0.5cm}

\section{INTRODUCTION}

In the present paper some issues of the low-energy neutrino physics
are considered that could be essential
for the current and future experiments aimed to search for a ``large''
neutrino magnetic moment. In the previous publication [1] we have studied
the inelastic weak and magnetic scattering of reactor antineutrinos
on the K- and L-shell electrons in a iodine atom. Here we extend our
calculations to the iodine M-shell electrons, and present the
results for the K- and L-shell electrons bound in a germanium atom. 
Besides the reactor $\bar\nu_e$'s, we also consider  
electron antineutrinos emitted by the $^{90}$Sr--$^{90}$Y and
$^{147}$Pm artificial sources. The use of artificial $\bar\nu_e$ sources
and semiconductor germanium detectors in experiments on the neutrino
magnetic moment has been considered in a number of recent publications
[2, 3](and references therein).

For the reactor $\bar\nu_e$'s, as was shown in ref. [1], the free
scattering differential cross sections can be approximately converted
to the inelastic ones by means of a simple step-function transformation.
Here we investigate the accuracy of such a recipe in more detail,
with different $\bar\nu_e$ energy spectra.

As an input, we use the standard power reactor $\bar\nu_e$
spectrum corrected for antineutrinos from the beta-emitters born 
in a core through the (n,$\gamma$) reactions [4]; for the 
$^{90}$Sr--$^{90}$Y and $^{147}$Pm sources we use the spectra tabulated
in ref.~[5]. The chosen input spectra are plotted in Fig.~1. The calculated
recoil electron energy spectra with these three sources
for the case of a free $\bar\nu_ee^-$-scattering are presented in Fig.~2.
In these calculations, eqs.~(2)--(4) of ref. [1] have been used.
The free magnetic- and weak-scattering recoil spectra are denoted as
$S_{\rm free}^M$ and $S_{\rm free}^W$, respectively. All calculations
in the present paper are done for the neutrino magnetic moment
$\mu = 2\cdot10^{-11}\mu_{\rm B}$ ($\mu_{\rm B}$ is the Bohr magneton), 
the constants of the electroweak interaction are the same as in
ref.~[1]. 

\section{INELASTIC SCATTERING ON ATOMIC \\
ELECTRONS}

In this section we consider the $\bar\nu_ee^-$scattering on electrons
bound in the iodine and germanium atoms. The energies and wave functions
of the discrete single-electron states are calculated within the relativistic
self-consistent Hart-ree-Fock-Dirac (HF-D) approach, with a local
exchange-correlation potential. The wave functions of outgoing electrons
in the continuum were obtained by a numerical integration of the Dirac
equation in the HF-D mean field. The details of this approach can be found
in refs. [6]. The calculated energies of some electronic subshells are
listed in Table~1, they agree with spectroscopic data within a few percent.

\begin{table}[t]
\begin{description}
\item[Table 1.]
\begin{center}
Calculated HF-D binding energies (in keV) of some\\
electronic shells in iodine and germanium atoms
\end{center}
\end{description}
$$\mbox{%
\begin{tabular}{c|c|cccc|cccccc}
\hline
&\multicolumn{11}{c}{ }\\
&\multicolumn{11}{c}{Shell}\\
&\multicolumn{11}{c}{ }\\
\cline{2-12}
&&&&&&&&&&&\\
Atom &K& &L$_{\rm_I}$ & L$_{\rm II}$ & L$_{\rm III}$ && M$_{\rm_I}$ &
M$_{\rm II}$ & M$_{\rm III}$ & M$_{\rm IV}$ & M$_{\rm V}$\\
&&&&&&&&&&&\\
\cline{2-6} \cline{7-12}
&&&&&&&&&&&\\
&1s$_{1/2}$&&2s$_{1/2}$&2p$_{1/2}$&2p$_{3/2}$&&3s$_{1/2}$&
3p$_{1/2}$&3p$_{3/2}$&3d$_{3/2}$&3d$_{5/2}$ \\
&&&&&&&&&&&\\
\hline
&&&&&&&&&&&\\
I &32.9&&5.09&4.78&4.48&&1.03&0.90&0.84&0.61&0.60\\
&&&&&&&&&&&\\
Ge&10.9&&1.35&1.22&1.18&&0.18&&&&\\
&&&&&&&&&&&\\
\hline
\end{tabular}}$$
\end{table}

For an iodine atom, the weak and magnetic inelastic 
$\bar\nu_ee^-$-scattering cross sections have been calculated for K-,
L- and M-shells which contain in total 28 electrons, the
remaining 25 electrons with binding energies less than 200 eV were 
considered as free. For germanium, atomic binding of K- and L-shell
electrons has been taken into account, all the other electrons were
treated as free. We shall denote the inelastic recoil energy spectra
for electrons knocked out from an atom due to the magnetic and weak 
interactions by $S^M_{\rm in}$ and $S^W_{\rm in}$, respectively. 

Shown in the top panels of Figs.~3 and 4 are the results of 
calculations of the recoil spectra $S^M_{\rm in}$ and $S^W_{\rm in}$,
respectively, for electrons knocked out from different electronic
subshells of an iodine atom by $\bar\nu_e$'s from the $^{90}$Sr--$^{90}$Y
source. In the bottom panels in these figures, the ratios to the
corresponding free recoil spectra are plotted. It can be seen that the 
magnetic scattering is strongly suppressed by the atomic binding effect
for the K- and L-shell electrons, and even for the M-shell with lower
binding energy, the effect of suppression is quite noticeable. Atomic
electron binding also influences the weak scattering cross sections but
the effect is much weaker. This unfavours the detectability of the
neutrino magnetic moment.

As follows from our calculations, the ratios 
$S^{M,W}_{\rm in}(T)/S^{M,W}_{\rm free}(T)$ for the other sources and target 
atoms are typically of the same character as those presented in
Figs.~3 and 4.

Let us introduce now, instead of the electron kinetic energy $T$,
the energy transfer $q$ which is defined by
\be
q =\Delta E =  {\epsilon}_{i} + T\,,
\ee
where $\Delta E$ is the neutrino energy loss in the inelastic scattering
process, $\epsilon_i$ is the electron binding energy on the
considered atomic shell (the nuclear recoil is neglected). For the
scattering on free electrons one has $\epsilon =0$ and $q=T$. In practice,
for majority of detectors, $q$ is just the total energy that could be
recorded as a visible energy of the event since the soft X-rays and Auger
electrons, ejected to fill the vacancy in the shell, are absorbed in the
detector sensitive volume and their summed energy, i.e. $\epsilon_i$, is
added to the kinetic energy $T$ of the recoil electron.

It has been demonstrated in ref. [1] that, for reactor antineutrinos,
the inelastic scattering spectrum $S^i_{\rm in}(q)$ for electrons from
the subshell $i$ of the iodine atom can be obtained from the free scattering
spectrum $S_{\rm free}(q)$, taken at the same visible energy $q$, if one
introduces the response function $R^i(q,\epsilon_i)$ defined by the relation
\be
 S^i_{\rm in}(q) = R^i(q,\epsilon_i) \times S_{\rm free}(q)\,,
\ee
in which $R^i(q,\epsilon_i)$ is approximated by the Heaviside 
step-function $\theta(q-\epsilon_i)$:
\be
R^i(q,\epsilon_i)\approx\theta(q-\epsilon_i) = \left\{
\begin{array}{ll}
 1, & \ q \ > \ \epsilon_i \\
 0, & \ q \ < \ \epsilon_i.
\end{array}
\right.
\ee

In the following, this will be referred to as the zero (or step-function)
approximation. 

Now, for each antineutrino source under consideration, we calculate
the ``exact'' spectra $S^i_{\rm in}(q)$ of electrons knocked
out from the iodine and germanium atomic subshells in the magnetic
and weak $\bar\nu_ee^-$-scattering processes, evaluate the ``exact''
partial response functions 
$R^i(q,\epsilon_i) \equiv S^i_{\rm in}(q)/S^i_{\rm free}(q)$ and
analyze their deviations from the zero approximation of eq.~(3).
``Exact'' means that the calculations are performed with the HF-D model.
Some of the obtained results are discussed below. 

For the L- and M-shell electrons, as can be seen in Figs. 5 and 6, the
``exact'' response functions both for magnetic and weak inelastic
scattering processes differ only slightly one from the other. They,
however, are close to unity everywhere except the rather narrow regions
near the corresponding threshold energies $q=\epsilon_i$. The same is true,
in case of magnetic scattering, for the K-shell electrons; on the other
hand, the weak scattering K-shell response function exceeds unity at
sufficiently large energies. This latter effect reflects the influence of
the atomic binding at relativistic energies [7] due to which the weak
inelastic cross section could be enhanced by a factor of the order of 
$1+O(\alpha^2Z^2) \sim 1+O(\epsilon_K/mc^2)$ ($\alpha$ is the fine
structure constant, $m$ the electron mass, $\epsilon_K$ the K-shell
binding energy); such an effect might be of relative importance only
for the tightly bound K-shell electrons.

For the soft $^{147}$Pm antineutrinos, as seen in Figs.~7 and 8, the zero
approximation {\it does not} provide a good fit for the weak scattering on the
K-shell electrons either in iodine or germanium atom. We also note that the
step-function approximation is strongly violated at the energy transfers around
the free scattering upper limit $q^{\rm max} = 2E^2/(2E+mc^2)$ ($E$ is
the incoming $\bar\nu_e$ energy). For $q>q^{\rm max}$, the free scattering
cross sections are identically zero while the inelastic ones are small but
still finite in the ``kinema-tically forbidden'' region beyond
$q^{\rm max}$~[6,7]. This tendency can be clearly seen for all iodine and
germanium subshells: for the energy transfer $q>80$~keV, as demonstrated
in Figs.~7-9, the ratios $S^i_{\rm in}(q)/S^i_{\rm free}(q)$ rapidly
increase with $q$.

\section{HOW GOOD IS THE STEP-FUNCTION \\ APPROXIMATION?}

To obtain an ``exact'' visible-energy-transfer electron spectrum 
$S^Z_{\rm in}(q)$ for a given atom, we have to sum all the partial
contributions $S^i_{\rm in}(q)$ from different subshells:
\be
S^Z_{\rm in}(q)=\sum (n_i/Z)S^i_{\rm in}(q)=S_{\rm free}(q)\cdot \sum
(n_i/Z)R^i(q,\epsilon_i)
\ee
Here $n_i$ is the number of electrons in the $i$-th subshell; the
spectrum is normalized to one electron. According to Eq.~(4), the
``exact'' response function for the whole atom has the form
\be
 R^Z(q) \equiv S^Z_{\rm in}(q)/S_{\rm free}(q) = 
\sum (n_i/Z)R^i(q,\epsilon_i)\,,
\ee
and, correspondingly, the atomic response function in the zero
approximation reads 
\be
R^Z_0(q) = \sum (n_i/Z)\theta(q-\epsilon_i)\,.
\ee
One sees that $R^Z_0$ is defined only by a set of the shell energies 
$\epsilon_i$ and shell occupancies $n_i$ of the particular atom.
It is the same for magnetic and weak scattering and does not depend on
the $\bar\nu_e$ source. For the iodine and germanium atoms,
the functions $R^Z_0$ are shown in Fig.~10. 

To check the validity of the
step-function approximation for the three $\bar\nu_e$ sources considered
in the present paper, we have calculated the ``exact'' electron energy
spectra $S^Z_{\rm in}(q)$  for the weak and magnetic $\bar\nu_e$ scattering
on these two atoms. We have got in total 12 different spectra. Then 12
corresponding ``exact'' response functions~(5) have been evaluated and
com-pared with those obtained in the zero approximation of Eq.~(6). The main
output of this comparison can be summarized as follows:

{\it In the $q=$(1.5--100)~keV energy range for the reactor and
$^{90}$Sr--$^{90}$Y $\bar\nu_e$ sources, and in the $q=$(1.5--80)~keV
range for the $^{147}$Pm $\bar\nu_e$ source, the relative deviation
$\mid R^Z_0-R^Z\mid/R^Z_0$ does not exceed (1.5--2)\% except
the case of weak scattering of $^{147}$Pm $\bar\nu_e$'s on an iodine
atom where the deviation can reach (3--4)\%.}

\section*{CONCLUSIONS}

A simple prescription is formulated on how to take into account the
effects of atomic electron binding: To a good approximation, to obtain
a visible-energy single-electron spectrum for the atomic inelastic
$\bar\nu_ee^-$--scattering, it is sufficient to multiply 
the free-electron spectrum by the ``zero response function'' $R^Z_0 = \sum
(n_i/Z){\theta}(q-\epsilon_i)$. Such a step-function approximation can be
safely used in the majority of current and future experiments aimed at
searches for the neutrino magnetic moment. 

We note, however, that this prescription is not universal. It can be used
for some specific though practically important cases like those considered 
in the present paper. For much softer antineutrinos (e.g., from tritium
source), the direct accurate calculations of the cross sections for 
$\bar\nu_ee^-$--scattering on the atomic electrons are needed.

\section*{ACKNOWLEDGMENTS}
This work is supported in part by the Russian Foundation for Basic
Research (RFBR), project 00-15-96708.

\end{document}